\documentclass[a4paper,11pt]{article}
\pdfoutput=1 % if your are submitting a pdflatex (i.e. if you have
\usepackage{jcappub} % for details on the use of the package, please
\usepackage[T1]{fontenc} % if needed
\usepackage[utf8]{inputenc}

\title{\boldmath The electric charge of black holes within galaxies}

\author[a]{Nelson D. Padilla,}
\author[b,c,d]{Ignacio J. Araya}
\author[a,e]{and Federico Stasyszyn}

\affiliation[a]{Instituto de Astronomía Teórica y Experimental, IATE, CONICET-UNC, Laprida 854, X5000BGR, Córdoba, Argentina}
\affiliation[b]{Instituto de Ciencias Exactas y Naturales, Universidad Arturo Prat, Playa Brava 3256, 1111346, Iquique, Chile}
\affiliation[c]{Facultad de Ciencias, Universidad Arturo Prat, Avenida Arturo Prat Chac\'on 2120, 1110939, Iquique, Chile}
\affiliation[d]{
School of Mathematics and Hamilton Mathematics Institute, Trinity College, Dublin 2, Ireland}
\affiliation[e]{Observatorio Astronómico de la Universidad Nacional de Córdoba, Laprida 854, X5000BGR, Córdoba, Argentina}

\emailAdd{nelson.padilla@unc.edu.ar}

\abstract{
We present improved estimates of the electric charge that black holes could hold when these are embedded in the ionised plasma within {galaxies}.  We have implemented the spontaneous emission of charges of opposite sign to that of the black hole via athermal Hawking evaporation, including its dependence on black hole spin, and we have estimated the equilibrium charge that arises as this charge loss is balanced by the continuous accretion of charges from the surrounding plasma.  The resulting charge can be several orders of magnitude lower than previously estimated upper limits, but it can surpass the pair production limit noted by Gibbons (1974) by a margin that increases with the amplitude of the black hole spin and the density of the plasma. We also implement a calculation for the net charge of {galaxies due to their stellar black holes and to primordial black holes in the case these make up a fraction of the dark matter. We find that these charges lie within the range that would produce adequate magnetic field seeds at the onset of galaxy formation.}
}

\begin{document}
\maketitle
\flushbottom

\section{Introduction}

%Our current picture of the Universe contains several unsolved mysteries, such as the existence and nature of the Dark Matter (DM) and the origin of magnetic fields.  

%The evidence for the existence of dark matter is quite varied but indirect, including early studies of the rotation curves of spiral galaxies \cite{Freeman70,Rubin1978}, and more recent results from the large-scale structure of the universe \cite{alam,eboss} and cosmological parameter fits to the temperature fluctuation power spectrum of the cosmic microwave background \cite{Planck16,Planck:2018vyg}, among other probes.

%Regarding the possible nature of dark matter, 

{Black holes are commonly thought to inhabit galaxies specially after the first detections of X-ray binaries which are accepted to host a black hole accreting material from its companion (e.g. \cite{Tanaka, macleod}).  Black holes have also been detected in our Galaxy via microlensing \cite{Bennet,Mao}.  These black holes are expected to form as the result of the end of the life of massive stars and consequently their masses range between $3$ and $100M_\odot$; their contribution to the total mass of galaxies should be at the subpercent level.  On the quite higher mass range there is ample evidence for the existence of supermassive black holes at the centers of galaxies, which are thought to span masses of up to billions of solar masses \cite{Cattaneo}.  These black holes are expected to lie in the centers of galaxies, with the exception of possible central black holes of accreted galaxies which have not yet  merged with the one of the host galaxy.  In all, these black holes are also thought to contribute very little to the mass of galaxies.}  

{Black holes of masses both within and outside these broad ranges could, in principle, form very early in the Universe and are generically referred to as primordial black holes \cite{Zeldovich1966,Hawking71,Hawking75}.  Depending on the mechanism for their formation, they can have from subsolar to supermassive black hole masses and higher still. There are primordial black hole (PBH) mass windows where they could even add up to that of the missing matter in galaxies and in the Universe, i.e. they could conform the dark matter \cite{Carr2020, sureda}. PBHs are possible candidates as they form early enough, considering that dark matter should be present since very early times \cite{Planck16,Planck:2018vyg,Arbey:2021gdg,Khlopov_2010}. }

There are several other compelling candidates for the dark matter apart from PBHs consisting of thermally produced fundamental particles that interact with other species mostly via gravity, as is the case of particles from extensions to the standard model of particle physics that produce the weakly interacting massive particles (WIMPS, \cite{Feng:2010,Arcadi:2018EPJC,wimps:2019}).  

The search for WIMPs in dark matter experiments has so far produced no direct detections \cite{Bernabei08,Mack07,Albuquerque10,Drukier86}, and the same is true of searches of effects due to black holes on different observables, which provide strong constraints on the fraction of dark matter that can be stored in black holes, which should be well below unity over most of the range of possible black hole masses, except in two  windows around the asteroid mass and the solar mass \cite{Carr:2020a, Carr:2020b,Sureda:2020,PadillaPkPoisson21}.  

{Black holes would produce additional observables to those related to the gravitational effect due to their mass.  For instance they are expected to emit radiation via  Hawking evaporation \cite{Hawking71}, which would be specially important in very low mass black holes.  WIMPs, on the other hand, would also produce emission due to their self annihilation in the dense cores of galaxies \cite{Araya2014}, albeit with a different signature to that expected from evaporating black holes.  }

In a recent paper, Araya et al. (2023, \cite{Araya23}, A23 from this point on) showed that  black holes could hold a non-zero electric charge.  {Their formalism can  be applied to black holes of any origin, including primordial and stellar ones.} %There are also particle candidates that could be electrically charged such as milli-charged particles \cite{Munoz:2018}.  The charge associated to  {matter}  provides additional observables. %that could serve to prove in another, independent indirect way, the existence of dark matter.  
In A23, the total charge of black holes is in most cases much smaller than the extremal value for the Reissneer-Nördstrom metric for charged black holes, meaning that the electromagnetic potential is always much smaller than that of gravity.  This places charged black holes below the current limit for charges for dark matter particles \cite{Caputo:2019}.  {Still, there can be additional observables that a charge would produce, opening the possibilities for their detection.}
The inferred black hole charge by A23 would depend on the black hole spin.

Even though there are multiple constraints on the abundance of primordial black holes as a function of their mass from observations, there are no constrains on the spin that these black holes could have.  Given that the charge of black holes could depend on their spin, it is interesting to review the literature for the current limits to its possible values.  Thorne (1974,\cite{thorne74}) investigated the possible spin-up of black holes as a result of accretion of material via a thin disc.  They found that once the spin is high enough, radiation from the thin disc itself effectively removes angular momentum from the black hole limiting the maximum dimensionless spin parameter to $a_{max}=0.998$.  Several authors agree with this estimate using different methods, also providing further extraction of angular momentum via the production of a jet \cite{meier01,nemmen07}.  However, there are  numerical estimates that allow the spin to reach much higher values such as $a=0.99997$ \cite{popham}.  This was confirmed later by Benson \& Babul \cite{benson}, but they argue that an evolving spin will probably reach its maximum at the Thorne limit.

{The initial spin of black holes of stellar origin is expected to be very low.  Candidates for stellar black holes accreting material from their companions in X-ray binaries allow their spin to be inferred from observations.  Since accretion is expected to increase the black hole spins \cite{macleod},  the measured spin from X-ray binary black holes is an evolved one.  On higher mass scales,} supermassive black holes do not seem to reach spins much higher than $a=0.6$ \cite{martinez}, but semi-analytic simulations of the co-evolution of supermassive black holes and galaxies in a cosmological context, show that the spin could reach values $a\sim 1$ \cite{lagos:2009}.  
These estimates of black hole spin are limited in precision which makes it difficult to estimate how close to the maximal one they actually are.

Opening the door for charged {matter in rotating galaxies} also brings along other possibilities, such as the production of magnetic fields.  Magnetic fields are ubiquitous in astrophysics \cite{Battaner97,Pandey15}, and are a necessary ingredient when trying to explain the  structure, dynamics, and evolution of the Universe across all scales \cite{Widrow02,Beck2013}. They exist in large and small structures, impacting particle acceleration and the evolution of the primordial plasma \cite{Kronberg02}. Studying magnetic fields at high redshift remains challenging, but their influence on plasma physics may soon be observable. Primordial magnetic field astronomy is a highly intense and challenging field in Cosmology, offering insights into the Early Universe and its impact on CMB anisotropies \cite{Barrow07,Subramanian06}. 

The origin, evolution, and amplification mechanisms of these fields are still unclear, leading to intriguing hypotheses \cite{Kandus11,Subramanian16,Yamasaki12,Barrow07}. Generating mechanisms often rely on a "seed field", raising questions about their creation, amplification, and large-scale distribution. Various approaches have provided answers consistent with observations, following the principles of Magnetohydrodynamics. Magnetic seed fields may have originated from the Big Bang or the Early Universe, possibly through battery-like processes under specific conditions \cite{Mtchedlidze2022,Durrive2015,Widrow02,Dar05}, or during phase transitions \cite{Quashnock89,Banerjee03,Durrer06,Saga15} (See the review by Widrow et al. for an overview of the different seeding mechanisms at different cosmic epochs \cite{Widrow2012}). 

{There are processes that can provide the seed field at later times, during the epoch of galaxy formation.  This is the case of spontaneous magnetic field generation by turbulent motions of the plasma \cite{zhou}, and proposals such as is presented Araya et al. (2020, \cite{araya}, A20 from this point on) who show} that primordial black holes with extended Press-Schechter like mass functions are in some cases able to provide the required seeds if black holes acquire a magnetic monopole charge as they form \cite{Maldacena}. {In a companion letter \cite{Padilla23} we show that the required seeds can also be produced by imbalances in the charged carriers of different sign at the epoch of galaxy formation.}  {There are also proposals for generation of seeds from battery mechanisms during the formation of Population III stars \cite{Xu2008} that provide stronger initial magnetic fields.}

{  
Vazza et al. (2017,\cite{Vazza2017}) summarize several cosmological magnetic field seeding scenarios (i.e. an uniform seed, a seed following the density perturbations, a seed that approximates the turbulent dynamo, and an astrophysical one from stellar sources).
They find that, at z = 0, all seeds reach the galaxy cluster magnetization. However, there are large differences in the magnetic field structure on scales of
galaxies, filaments and voids, depending on the seeding mechanism. This emphasizes that the proposal of additional seeding mechanisms should also provide observable features.}

In this paper we make an extension of the formalism of A23 to explore the late time charge in black holes embedded in a plasma, to make a more detailed mapping of the charges and their dependence on black hole parameters.  A23 present two extreme cases of spin, namely maximally spinning ones with dimensionless spin parameter $a=1$, which {in the case of black holes of primordial origin} could preserve their initial charge, and black holes with zero spin.  Here we explore the intermediate cases as there is ample evidence that black holes that accrete matter, even if their initial spin is null, can spin up to $a\sim 1$ \cite{thorne74}.  We find that black holes with higher spin can hold higher electrical charges.  {We also estimate the total charge in black holes in galaxies, and compare this to the charge required for a rotating proto-galaxy to produce a seed magnetic field of the right amplitude.}

%We then explore whether the charge in these black holes combined, in some cases, with the induced charge imbalance in the baryonic plasma in rotating proto-galaxies is able to produce the required seeds for the cosmic magnetic fields without resorting to magnetic monopole charges as in A20.  We also look into the possibility that  black holes of astrophysical origin can acquire electrical charge from the baryonic plasma of the first galaxies and can therefore provide the required seed field.  

\subsection{Outline and conventions}

This paper is organised as follows.  Section \ref{sec:charge} expands the model for the charge accretion and emission by black holes in an ionised plasma within galaxies, it looks into the timescales involved in these two processes and  presents an equilibrium charge.  Section \ref{sec:halocharge} presents {the total galaxy charge and its density profile. We also compare the total galaxy charge in black holes to the one needed to explain the cosmic magnetic field seeds.}  
We present a discussion of our results and our assumptions in Section \ref{sec:discussion}, and conclude in Section \ref{sec:conclusions}.

For our calculations we adopt a flat cosmology with $\Omega_{m,0} = 0.315$; $\Omega_{dm,0} = 0.264$; $\Omega_{r,0} = 9.237 \times 10^{-5}$ and $H_0 = 67\, \frac{\mathrm{km}}{\mathrm{s}} \mathrm{Mpc}^{-1}$, consistent with the latest measurements from the Planck Satellite \citep{Planck:2018vyg}.  When considering that a fraction of the dark matter is in primordial black holes, we assume that they all have equal mass, i.e. a monochromatic primordial black hole mass function.  Throughout we assume that galaxies are hosted by dark matter haloes %(sometimes referred to simply as haloes), 
and will refer to the total mass of galaxies including their dark matter haloes as the virial mass.  {We will frequently focus on Milky Way like galaxies since these are the typical galaxies of the Universe from the point of view of their stellar populations (e.g. \cite{baldry})}.   We adopt the international system of units, and use the following notation for constants where $e$ is the absolute value of the electron's charge, $c$ stands for the speed of light, $k_c$ is Coulomb's constant, $G$ is the universal gravity constant, $h$ is Planck's constant and $\hbar$ its reduced version, $m_p$ and $m_e$ are the proton and electron masses, $k_B$ is Boltzmann's constant and $\mu_0$ is the magnetic permeability in vacuum.

\section{Charge of  black holes within galaxies}
\label{sec:charge}

{We will consider black holes of two distinct origins.  First, due to the ample evidence of the existence of black holes resulting from the death of massive stars \cite{mirabel,macleod}, we will consider stellar black holes. Second, we will also consider black holes of primordial origin formed close to the end of Inflation.  As we will discuss in Section \ref{sec:discussion} our formalism is not applicable to supermassive black holes.}

\subsection{Initial charge}

%We will consider that black holes can hold a charge that can come from two distinct origins; that it was acquired at the formation event of the black hole and that it lasted unperturbed to the present day, or that they {formed with zero charge and acquired it} at late times.  
%We will consider black holes of primordial and stellar origin indistinctly both of which could be present in galaxies. 

{Stellar black holes  will be assumed to hold no initial charge. They will be considered to hold a range of spins taking into account their likely zero spin at formation and their possible spin up in X-ray binaries.}

{We will allow primordial ones to contain a non-zero initial electric charge and a dimensionless spin parameter that can range from zero to maximal.  We will consider two different types of primordial black holes}.  Ones that could form from overdensities left over from Inflation, that as they enter the horizon early during  radiation domination they collapse and form a black hole that could hold a non-zero charge due to the unrelaxed spatial distribution of positive and negative charges at the horizon scale that is just coming into causal contact. In A23 this charge is referred to as the Poisson initial charge.  For a Lagrangian volume, $V_0$, that just enters the horizon the charge is,
\begin{equation}\label{eq:Q0poisson}
Q_0 = \pm \sum_i |q_i|\sqrt{V_0n_i},
\end{equation}
where the index $i$ runs over all the particles present at temperature $\bar{T}_0$, $n_i$ is the number density of each particle, 
\begin{equation}
V_0 = \frac{M c^2}{\rho_{\text{\tiny{uni}}}(\bar{T}_0)},
\end{equation}
$M$ if the mass of the formed black hole, and $\rho_{\text{\tiny{uni}}}(\bar{T}_0)$ is the average energy density of the Universe at $\bar{T}_0$. 

The second type of primordial black holes could be formed by the high energy collision of equal charge fundamental particles that could produce extremal Reissner-Nördstrom (RN) black holes of at least one Planck mass, with a charge that ensures extremality,
\begin{equation}\label{qmax-saulo}
    Q_{\text{\tiny{max}}}\sim\sqrt{\frac{G}{k_c}}m_{\rm Planck},
\end{equation}
where $m_{\rm Planck}$ is the Planck mass.  This charge could be achieved by the collision of two charged particles taking into account the running of the electric charge with the energy scale of the
collision \cite{Peskin95}.

In A23 it is proposed that if the dimensionless spin parameter is maximal, $a=1$ then the initial charge could be held within, or in close proximity of, the black hole such that for the distant observer the charge would be seen as remaining constant all the way to the present day.  

\subsection{Charge evolution}

{Here we concentrate on late-time charge evolution within galaxies.  Therefore, a main ingredient for the charge evolution is the galaxy plasma in which black holes are embedded.  In what follows we will consider a simplified view of the barionic gas within galaxies, assuming it is a two-phase medium as is usually done in semi-analytic models \cite{baugh} consisting of a hot phase that satisfies the conditions of a plasma, and a cold phase that contains the gas that cooled radiatively from the hot, plasma phase.  This is a simplified view that serves to make order of magnitude estimates of black hole charges and their time evolution.}

{Stellar black holes will be assumed to share their dynamics with stars in the galaxy, i.e. different from that of the ionised gas of the galaxy plasma.  Black holes will likely spend a non-zero fraction of their orbits within the latter.}

{Primordial black holes that form soon after the Big Bang will be subject to charge evolution by acquiring charges from its surroundings down to the epoch of recombination.} For the case of non-maximal spin A23 show that the initial charge is lost by {this time}.  

However, they also show that later on when black holes {of stellar or primordial origin} find themselves within the ionised plasma of galaxies, they can acquire a charge since the late-time charge carriers with opposite charge sign have masses that differ by several orders of magnitude; i.e.  electrons are more likely to be accreted. As a result {stellar or} primordial black holes with non-maximal spin should acquire a late time electric charge.   

The following subsection explores the charge that such non-maximally spinning {stellar or primordial} black holes would be able to acquire as a function of black hole mass and spin.

\subsubsection{Charge accretion and emission of non-maximally spinning black holes in an ionised plasma}

In order to reach a reasonable estimate of the final black hole charge within a plasma in galaxies, A23 took into account processes of spontaneous charge production at the outer horizon of black holes, namely the Hawking radiation process \cite{Hawking75} and the Schwinger mechanism \cite{Schwinger,Gibbons75}.  For the latter they did not include a detailed calculation of the emission due to vacuum polarisation; instead they only checked whether the estimated charges were below the rapid discharge limit estimated by Gibbons (1975, \cite{Gibbons75}), which was satisfied by all their estimates.
For the Hawking evaporation, A23 only made the calculation for black hole (BH) masses $M\lesssim2\times10^{13}$kg which corresponds to the BH mass that satisfies $k_BT_{\rm H}\sim m_e c^2$, susceptible to charge emission by thermal Hawking radiation.  

In this paper we improve the estimates of Hawking evaporation charge emission in higher mass black holes using Hawking radiation which, even when the Hawking temperature is much lower than the rest mass of electrons, is able to emit the latter due to the chemical potential produced by the black hole charge, as postulated by Lehman et al. \cite{Lehmann}, who term this emission as athermal.  Furthermore, Lehmann et al. also estimate that the charge emitted via the Schwinger mechanism is of the same order of magnitude as the athermal Hawking emission, which allows us to use our Hawking athermal emission calculations as good measures of the total charge emission.

As in A23, we estimate the charge emission via Hawking radiation adopting the distribution function,
\begin{equation}
    f(\epsilon(p),\mu)=\frac{1}{\exp{\left(-\frac{\epsilon(p)-\mu}{k_B T_{\rm H}}\right)}+1},
\end{equation}
where $$T_{\rm H}=\frac{ \hbar}{ 2 \pi G k_B}\ \frac{ \sqrt{(1-a)M^2-\frac{k_c}G Q^2}}{2 M^2+2M\sqrt{(1-a)M^2-\frac{k_c}G Q^2}}$$   is the Hawking temperature of the BH that takes into account the spin and charge of the black hole \cite{Ruiz2019},   $\epsilon(p)=\sqrt{p^2c^2+m_e^2c^4}$ is the kinetic energy of a single particle with momentum $p$ and mass $m_{e}$\footnote{Note that the distribution is of the Fermi-Dirac type as indicated by the $+1$ in the denominator; the $-1$ in A23 is simply a typo present only in their manuscript and not in their calculations (Araya, private communication).}.  We introduce a modification in $\mu$, the chemical potential that accounts for the Coulomb energy of the particle with electric charge $\pm e$ sourced by the BH charge $Q$, by allowing it to include the magnetic field due to the spin of the black hole,
\begin{equation}
    \mu(\pm e)\simeq\frac{(1-a) k_cQ(\pm e)}{r_+},
\end{equation}
%\begin{equation}
%    \mu(\pm e)=k_cQe/r^+\simeq m_e c^2
%\end{equation}
%\begin{equation}
%    Q=\frac{m_e c^2 r^+}{k_c e}=\frac{2GM}{k_c e}=:Q_{\rm pairs}
%\end{equation}
%\begin{equation}
%\frac{e}{10^{-18}M_\odot}
%\end{equation}
%\begin{equation}
%Log_{10}({\rm Halo Mass}/h^{-1}M_\odot)
%\end{equation}
where we have made use of the simplification that the averaged amplitude of the dipolar magnetic field energy is simply that of the Coulomb one multiplied by the dimensionless spin parameter $a=cJ/(GM^2)$ (derived from \cite{Carter}) valid when $a=0$ or $a\sim1$, with $J$ the angular momentum of the black hole, and where $r_+$ is the outer horizon, 
\begin{equation}
    r_+=\frac{r_{\text{s}}+\sqrt{r_{\text{s}}^2-4r_{a}^2-4r_{Q}^2}}{2},
\end{equation} 
where $r_Q^2=Q^2 G/(4\pi \epsilon_0 c^4)$ and $r_a=J/(Mc)$. 
This simple assumption allows us to explore the effect of black hole spin on athermal Hawking charge emission.  {Note though that even when the spin is very close to the maximal one, i.e. for very small $(1-a)$ , the chemical potential is still non-zero and able to produce athermal discharge when $k_BT_{\rm H}\ll m_e c^2$, although at a slower rate.} 

We follow A23, and use the charge density $\rho$ of created electrons or positrons in the vicinity of the BH horizon to write the current $J=\rho v$ that can be used to obtain the rate of charged particle loss by the black hole,
\begin{equation}
    \dot Q_H(\pm e)=\int \vec J(\pm e)\cdot d\vec A=A\int \frac{2d^3p}{(2\pi \hbar)^3} v(p) \cos(\theta(\vec p, \hat n)) (\pm e) f(\epsilon(p),\mu(\pm e)).
\end{equation}
Adding together the emission of electrons and positrons, regardless of the BH charge sign, the final discharge rate due to Hawking emission and its dependence on black hole mass, charge and spin, reads
\begin{equation}
    -\dot Q_H(M,Q,a)= A\int_{0}^{\infty}\frac{dp p^2}{(2\pi)^2 \hbar^3} v(p) (- e)\left( f(\epsilon(p),\mu(-e))-f(\epsilon(p),\mu(e))\right).\label{Eq:Hawkdischarge}
\end{equation}
This approximation ignores several details such as grey body correction factors {of $\mathcal{O}(1)$}, but allows us to obtain order of magnitude estimates for the charge emission by massive black holes with non-zero spin.  Note that the inclusion of spin will also introduce changes to the final average charge of low mass black holes with Hawking temperature below the electron mass.

After implementing Eq. \ref{Eq:Hawkdischarge} we find that if the black hole charge is larger than the Gibbons limit for pair production charge \cite{Gibbons75},
\begin{equation}
    Q^{\rm pairs}=\frac{2GMm_e}{k_c e},
\end{equation} 
then the rate of discharge is positive. For $Q\leq Q^{\rm pairs}$, the rate is null, i.e. a charge below $Q^{\rm pairs}$ is stable if no charges are accreted onto the black hole. {In a companion letter \cite{Padilla23} we note that this charge corresponds to an excess of one charge every $\sim10^{39}$ baryons, which corresponds to the maximum charge that would be able to be maintained by the gravitational pull of any massive object.}

One can compare the rate of discharge to the rate of charge accretion from the plasma surrounding the black holes.  A23 estimated the maximum charge that a black hole would acquire until its repulsive force would impede further charge accretion as,
\begin{equation}
    Q_{\rm max}=\frac{k_B T_{\rm vir} \left( \frac{1}{r_+}-\frac{1}{r_++\ell_{\rm Debye}}\right)^{-1}+GMm_e}{k_c e}.
\end{equation}
where they consider that electrons can move much faster than protons, and freely within a Debye length of the plasma,
\begin{equation}
    \ell_{\rm Debye}=\left(\frac{k_B T_{\rm vir}}{4\pi k_c n_e e^2}\right)^{1/2},
\end{equation}
where $n_e$ is the average number density of electrons.  We only take into account the number density of electrons since regardless of virial temperature ($T_{\rm vir}\leq10^{10}$K) they have a higher thermal velocity than protons.

The rate of charge emission when the charge is $Q_{\rm max}$ is effectively zero, but if this charge is larger than $Q^{\rm pairs}$ then the charge will drop due to Hawking athermal emission.  This emission fueled drop will continue until the negative charge accretion equals the emission, or until the emission stops at $Q^{\rm pairs}$.  

We follow A23 and assume that the cross-section for neutral BH charge accretion is the geometrical one, {which is adequate for black holes with no charge; here we add the modification that if the black hole charge is already non-zero, instead of adopting the innermost stable circular orbit, we use the distance at which the Coulomb and gravitational forces cancel each other out, 
\begin{equation}
    r_E^i=\frac{-k_c Q (\pm e)+ G M m_p}{3 k_B T_{\rm vir}},
\end{equation}
where $i$ stands for protons or electrons ($p$ or $e$). We limit this scale to two extreme values.  First, for charges below the limit where particles are energetic enough to be able to cross the Coulomb barrier, i.e.
\begin{equation}
k_B T_{\rm vir}>(G M_{BH} m_e+k_C Q e)\left(\frac{1}{r_{\rm isco}+\ell_{\rm Debye}}-\frac{1}{r_{\rm isco}}\right),
\end{equation}
we limit the accretion area to have a radius of the innermost stable circular orbit, $r_{\rm ISCO}$ {since the resulting charges are low enough that electrons and protons are able to break the Coulomb barrier and reach the black hole from a $\ell_{\rm Debye}$ distance away}.  On the upper side we limit the radius of the cross section to one Debye length away from $r_{\rm ISCO}$, since beyond this separation the potential of the charged black hole is shielded by the plasma.  We ignore the magnetic dipole force as it drops with an extra factor of $1/r$ compared to the Coloumb one. With this, the cross sections for protons and electrons read,
\begin{equation}
   \sigma_{\rm BH}^i(M)   =\begin{cases} \pi (1.4 r_+)^2&  1.4 r_+>r_E^i \text{ or } |Q|<e\\
      \pi (r^{i}_E)^2& 1.4 r_+<r_E^i<1.4 r_++\ell_{\rm Debye}\\
      \pi (1.4 r_++\ell_{\rm Debye})^2& r_E^i>1.4 r_++\ell_{\rm Debye}.
    \end{cases}
\end{equation}
 }
Then, the rate of charge acquisition (or loss) by BHs of mass $M$ reads \cite{longair1998galaxy},
\begin{equation}
    \Gamma(M)=n_e \left(\sigma^e_{\rm BH}(M)  v_e(T_{\rm vir})-\sigma^p_{\rm BH}(M)  v_p(T_{\rm vir})\right)+\dot Q_H(M,Q,a)/e,\label{eq:gammaH}
\end{equation}
where $v_e$ and $v_p$ are the electron and proton thermal velocity within the {galaxy}, and we have added the charge loss due to Hawking emission in the last term. Black hole stable charges that do not rise above $\sim Q^{\rm pairs}$ are not affected by Hawking athermal discharge.  

{Notice that the cross-sections for accretion of electrons and protons can be quite different depending on the sign of the charge of the black hole.  Even when electrons are thermally faster than protons, protons can have much larger cross sections for accretion, and in this case the initial negative charge acquired when the black hole was neutral, is rapidly lost, and possibly turns {positive}.  However, electrons can be rapidly acquired again, and the charge can be lost.  The case in which this does not happen is when the cross section radii are similar to $r_{\rm isco}+\ell_{\rm Debye}$. Depending on the plasma density, this condition is met for black holes with masses $M>10^{22}$kg, and black holes with larger masses can in general hold a charge as long as they are embedded in a plasma (or do not encounter free charges).}

We explore values of $n_e$ that correspond to matter densities ranging from $1$ to $10^{6}$ times the virial one defined as $200$ times the critical density, i.e., $n_e^{\rm vir}= 200 \Omega_b \rho_{c}/(1.4m_p)$ approximating the average ion nuclei weight to $1.4 m_p$ taking into account the presence of Helium. 

We integrate Equation \ref{eq:gammaH} in time using time intervals that do not produce changes in the charge larger than $1\%$.  We integrate up to a time such that the charge shows no further significant changes (if the rate is always of the same sign) or until an oscilatory behaviour is reached (when the rate alternates sign).  We refer to this final charge as the equilibrium  charge $Q_{eq}$.  Note that higher baryon densities produce higher accretion rates and increase the resulting equilibrium value. 

\subsubsection{Charging timescales of non-maximally spinning black holes embedded in a galaxy}

\begin{figure}
    \centering
    \includegraphics[scale=.5]{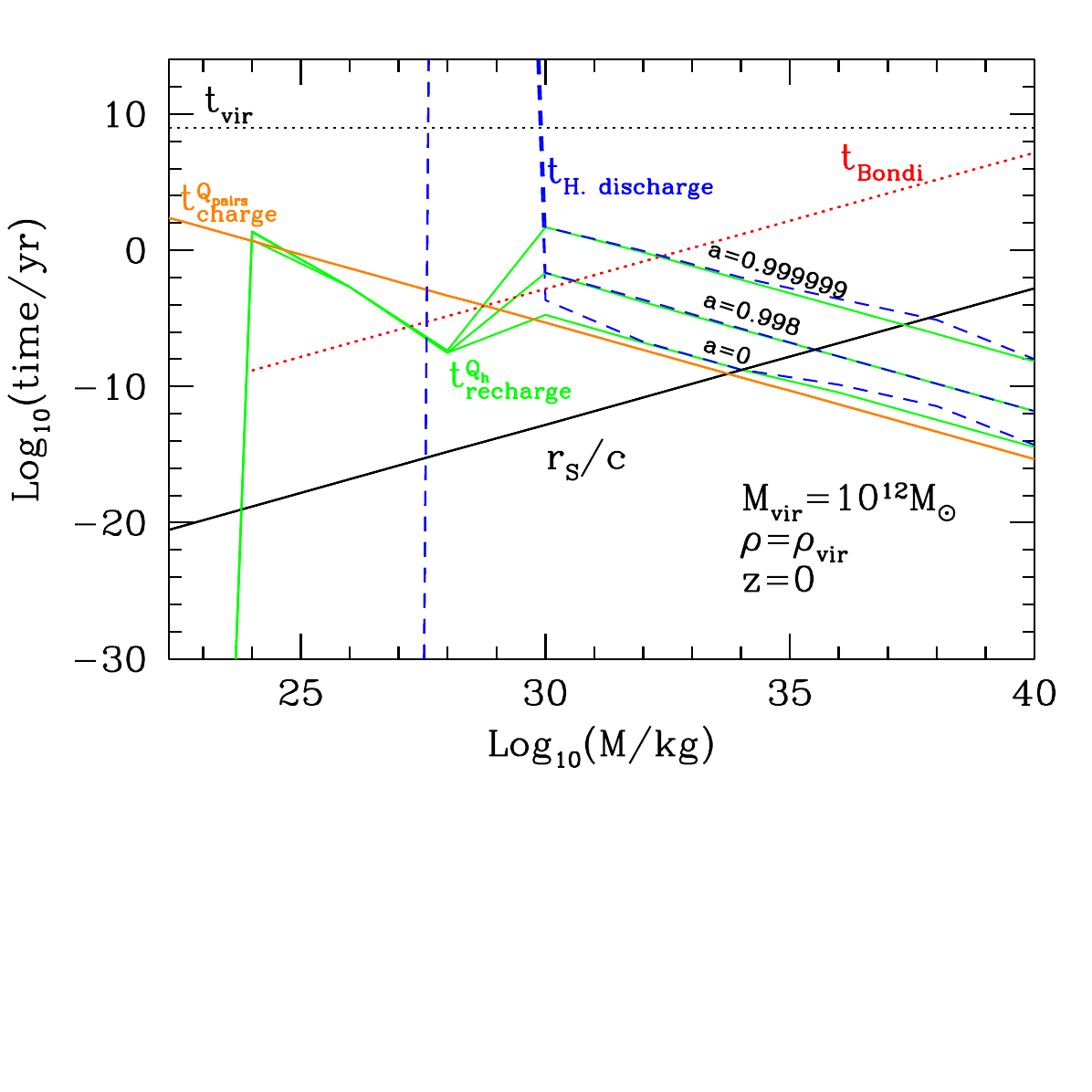}
    \vskip -3cm
    \includegraphics[scale=.5]{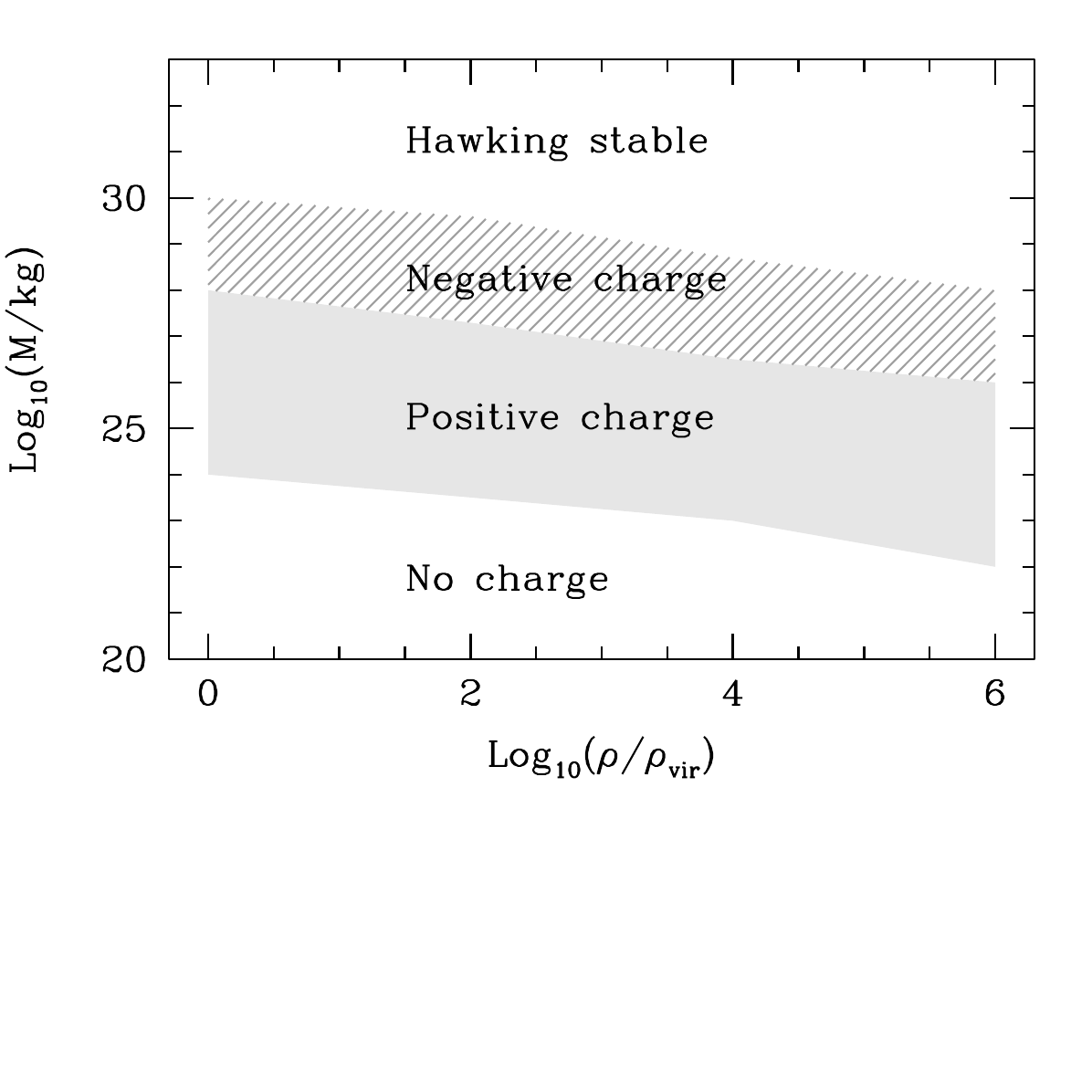}
    \vskip -3cm
    \caption{Top: timescales of charge accretion and emission by black holes.  The blue lines show charge emission due to Hawking radiation, for increasing values of the dimensionless spin parameter $a$ as shown in the figure key (larger timescales for larger spin).  Green lines show the recharge time for the equilibrium or stable charge.  Notice that at high masses the green and blue lines match closely, indicating that the equilibrium charge has been reached; in these cases the charge is $Q>Q^{\rm pairs}$.  When the green line lies below the blue one, the charge is rapidly acquired and an equilibrium between electron and proton accretion is reached.  The orange line shows the time required to acquire the limit of pair production discharge $Q_{pairs}$.  The horizontal dotted line shows the typical virial time of a $10^{12}$h$^{-1}M_{\odot}$ {galaxy}, and the violet dotted line shows the free-fall timescale for the Bondi mass, for BH masses where mass accretion from a plasma is possible. The solid black line shows the light travel time to traverse the Schwarschild radius of the black hole showing that up to $M\sim 10^{35}$kg the instantaneous Coulomb potential approximation is valid.  
    Bottom: ranges of black hole mass for Hawking stable (equal Hawking discharge and charge accretion times), negative stable charge (accretion of electrons is balanced by short bursts of accretion of protons, maintaining a stable charge), positive stable charge, and zero charge of $Q=e$, as a function of plasma density.  }
    \label{fig:timescales}
\end{figure}

We calculate the $z=0$ timescales of accretion and emission, $Q/\dot Q$, and show estimates as a function of black hole mass for different dimensionless spin parameters $a<1$, and show the results in the top panel of Figure \ref{fig:timescales}.  As can be seen, a larger dimensionless spin parameter results in a slower Hawking discharge above solar masses (blue dashed).  We specifically show the case with $a=0.998$ corresponding to the Thorne limit \cite{thorne74} for illustrative purposes. The green solid lines show that the accretion times are lower than discharge ones for masses below $\sim 10^{30}kg$, which implies that such black holes acquire their $Q_{\rm max}$ charge relatively fast and are able to keep it for long {(modulo exchange of protons and electrons to maintain equilibrium)}, sometimes even for longer than the dynamical time of a typical {galaxy} ($t_{\rm vir}$, shown as a horizontal dotted line); the charges of such black holes can be felt by a distant observer as soon as the black hole moves away from the region of the plasma where it acquired its charge.  {In some cases the black hole could take along a trailing group of protons that would effectively neutralise its charge (see Section \ref{sec:discussion}).} {Notice that these timescales are short, which implies that these results remain almost unchanged up to $z=20$, a redshift we will concentrate on later in this paper.}

We illustrate the black hole mass ranges that satisfy these conditions in the bottom panel of Fig. \ref{fig:timescales}, {which shows in grey and hatched regions the  black hole masses that can hold a charge for longer than the dynamical time of the {galaxy} (negative and positive charges, respectively)}, i.e. that can exert a Coulomb force that a distant observer can detect with its full charge, as a function of plasma density.  
Above this mass black holes reach an equilibrium where the charging and Hawking discharge times are equal (to within the numerical accuracy of our calculations).  In this case the black hole charge is stable, but charges constantly enter and exit the black hole at very high rates.  This charge is probably not measurable with its full amplitude by distant observers as there is a trail of exchanged charges left behind as the black hole  traverses the plasma.  This trail should rapidly relax back with the plasma  possibly leaving a thermal contribution to its temperature.  However, even if their stable charge is larger, these black holes can in principle be detected as having $Q^{\rm pairs}(M)$ as further discussed below.  The top panel of this figure shows the time required to obtain this latter charge as an orange line.  As can be seen, this charge is acquired quite rapidly in comparison to the virial time. %But in any case the spatial distribution of charges should be left unchanged by this process.   What a distant observer could detect though, is the last charge the black hole acquired before starting to discharge via athermal Hawking radiation, i.e. a charge of $Q^{\rm pairs}(M)$, even when its actual charge could be larger.  

The extent to which there are macroscopic, galaxy wide consequences of the charge of black holes can be tackled bearing in mind that the dynamics of a pressureless fluid composed of black holes and that of the plasma can be thought of as being independent.  If a black hole retains its charge for a long timescale and it does not capture from the plasma a trailing group of particles with opposite charge sign that could neutralise the BH charge, its effect will be seen by a distant observer (say, at least hundreds of parsecs away, in a different component of the {galaxy}, or from outside the {galaxy}).  %We will come back to this issue later in the Discussion section.  
On the other hand, if the charge is stable but a result of a continuous, almost instantaneous accretion and emission of charges (in terms of the dynamical time of the {galaxy}), then one possibility is that the net charge of the plasma and black holes together will be seen as zero throughout the galaxy by an external observer.  A second possibility is that, even though the net stable charge could be larger, since the initial stages of charging of a black hole up to $Q^{\rm pairs}(M)$ were not subject to emission discharges, the distant observer will be able to detect $Q^{\rm pairs}(M)$.  The black hole charge in this case would have been acquired from the plasma quite rapidly at some early time, and moved away from the gas that provided the electrons, effectively imprinting a difference in the spatial distribution of charges in black holes and in the gas within a galaxy.

Notice that all black holes are in principle subject to these charging and emission discharge processes. {This includes black holes}  of astrophysical origin which can make up to a significant percentage of the total stellar mass formed within galaxies and can also be considered to have independent dynamics to that of the baryonic plasma (see for instance \cite{Lacey16,Lagos:2011}); {and it also includes primordial black holes which, depending on their mass, can make a significant fraction of the dark matter and should also have independent dynamics than the plasma}.

\subsubsection{Instantaneous Coulomb potential approximation}

{The present} formalism assumes that the Coulomb potential produced by the black hole is instantaneously adjusted as individual charges are accreted or emitted by the black hole.  We check that this assumption is valid by adding in the top panel of Fig. \ref{fig:timescales} the light travel time for the Schwarschild radius (black).  As can be seen up to masses well above the solar mass, the light travel times are below the charging and discharge timescales for most of the black hole masses we will explore here.  This ensures that there is enough time for the potential to relax between  accretion or emission events. 

\subsection{Black hole charges within galaxies}

\begin{figure}
    \centering
    \includegraphics[scale=.5]{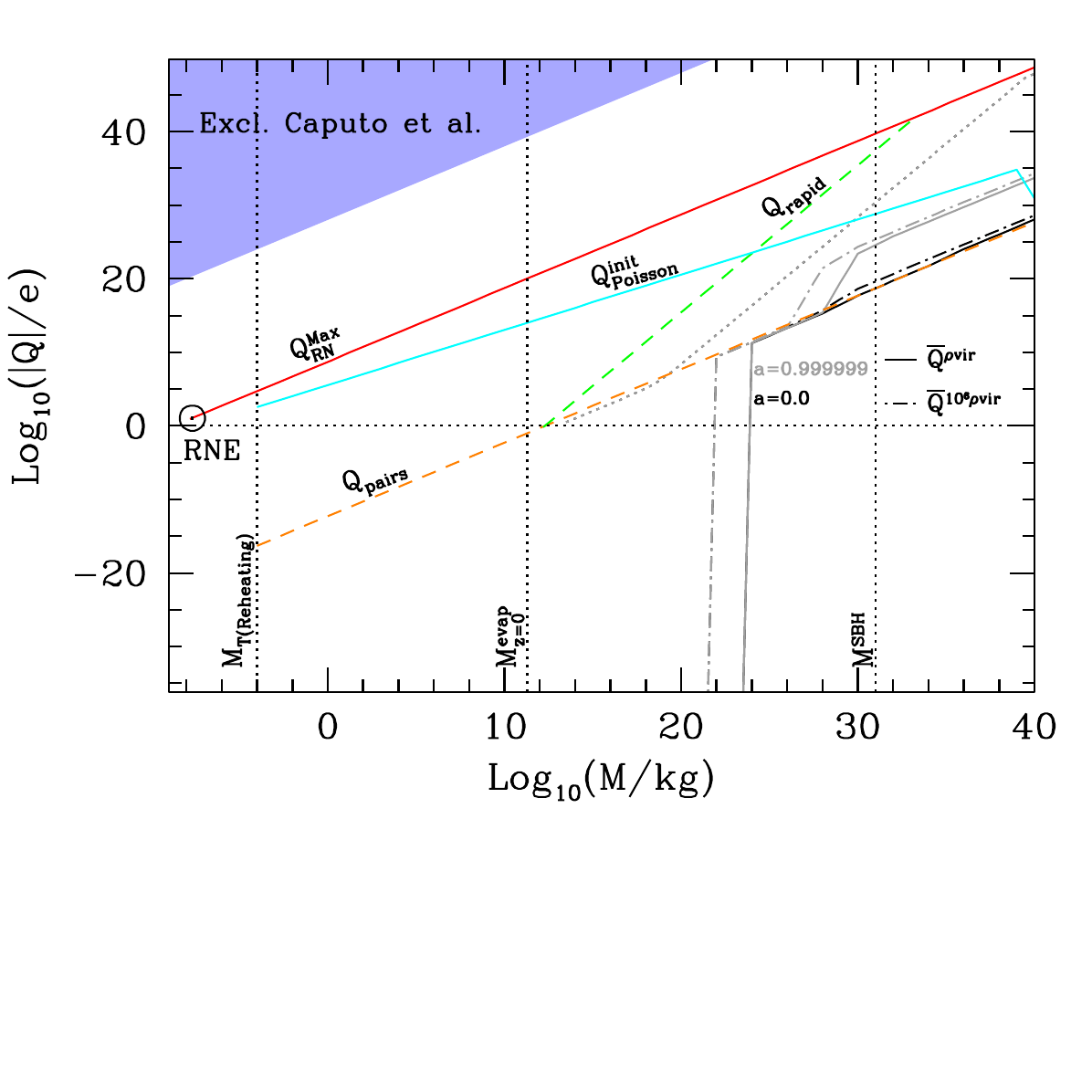}
    \vskip -3cm
    \includegraphics[scale=.5]{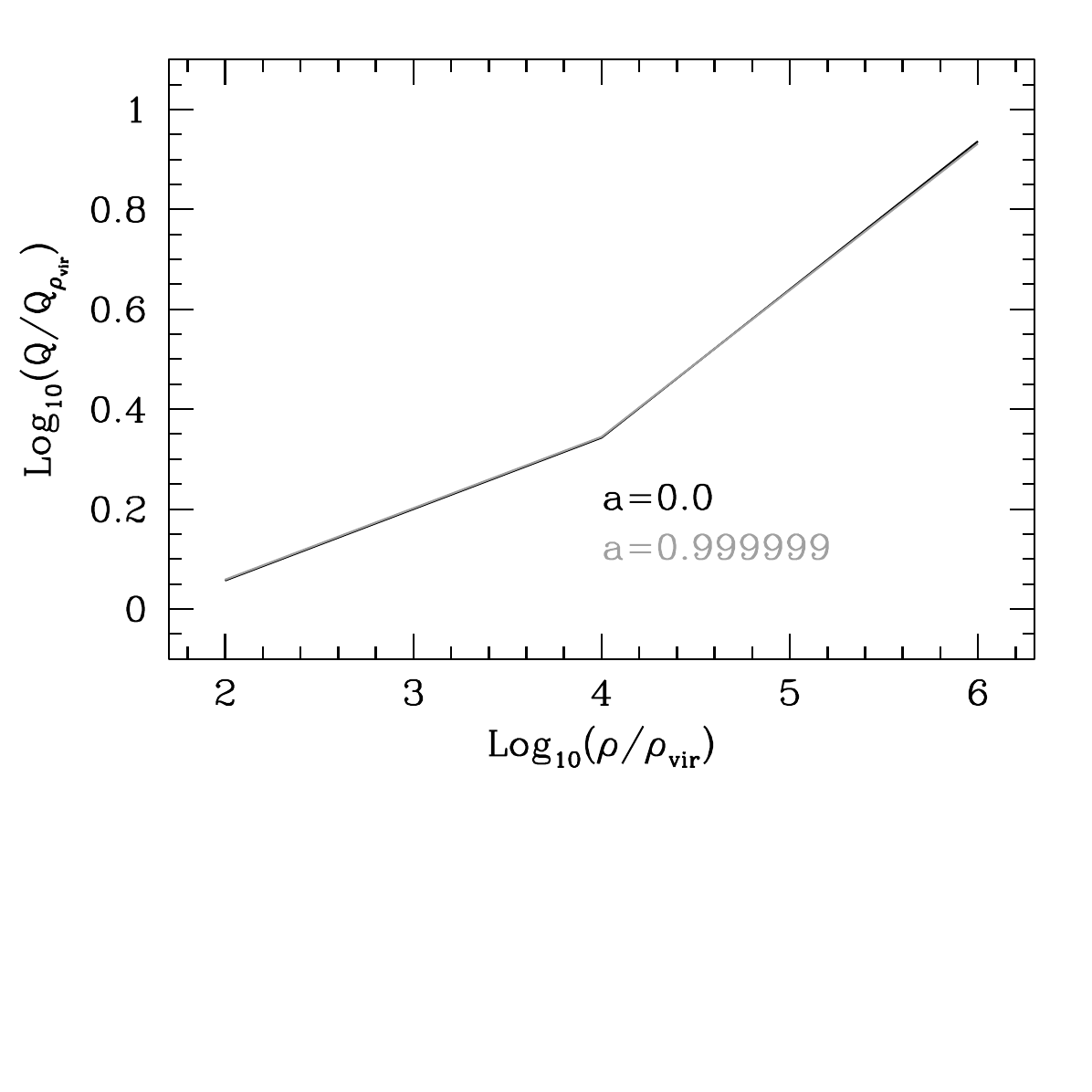}
    \vskip -3cm
    \caption{Top: black hole charge in units of $e$ as a function of its mass.  The cyan line shows the initial Poisson charge for primordial black holes, the red line the maximum charge for a RN extremal black hole. The circle shows the RN extremal charge for Planck mass primordial black holes. The dashed orange line shows the  charge for pair production by Schwinger and/or Hawking (athermal) radiation.  The dashed light green  line shows the rapid discharge limit of \cite{Gibbons75}.   
    The horizontal dotted line shows $Q=e$, and the vertical dotted lines show the black hole mass formed at reheating (left)  the mass of a primordial black hole that would evaporate by today (middle), and the approximate mass of black holes of stellar origin (right). {The  solid and dot-dashed lines show the stellar or primordial black hole charge acquired from  ionised baryonic plasmas with  virial density and $10^6$ times the virial density, respectively. Black lines correspond to a dimensionless spin parameter of $a=0$ and grey lines show the same for $a=0.999999$.}  
    Bottom: Ratio of the black hole charge embedded in a plasma as a function of density  in units of the virial one for $a=0$ and $0.999999$ (black and grey, lines overlap).}
    \label{fig:charge}
\end{figure}
The resulting $z=0$ final charges for non-maximally spinning black holes {as a function of black hole mass} are shown in the top panel of Figure \ref{fig:charge} as solid and dot-dashed black and grey lines, for different baryon densities and different dimensionless spin parameters that span a range that brackets the Thorne limit.  The baryon densities and spins are presented in the figure key and in the caption.  {The black hole charge does not depend appreciably on the galaxy virial mass for non-maximally spinning black holes with masses $M>10^{22}kg$.} The black dotted diagonal line shows the A23 estimate for the maximum charge a non-maximally spinning BH could acquire from the {galaxy} plasma\footnote{{We have chosen not to consider the case when a fraction of black holes hold a charge $Q=e$ presented in A23 simply because it would not be able to produce a galaxy-wide magnetic field.}}. The figure follows the style of A23 showing for reference the Schwinger charges for pair production and rapid discharge  \cite{Gibbons75} in orange and light green dashed lines, the RN extremal charge (red), the electron charge, minimum mass for primordial black holes, evaporated mass by today and stellar mass renmants (dotted horizontal, and left, centre and right vertical lines, respectively).  The black circle and cyan lines show the charge for maximally spinning {primordial} black holes for the extremal RN and Poisson initial charge cases, respectively.  %The grey region indicates that even if the spin of black holes is maximal, their charge could still be lost at some rate, which makes it possible for the charge to lie between the cyan line/circle of maximally spinning black holes and the late charge that non-maximally spinning ones can acquire within galaxies/haloes. 

As can be seen, when the spin is non-maximal, as soon as the black hole mass is such that its charge rises a few orders of magnitude above the pair production charge (orange) the Hawking athermal emission kicks in and prevents the charge from increasing further for higher black hole masses, which makes the final equilibrium charge to depart from the A23 maximum charge.  %Black holes in  halos/galaxies of lower mass acquire lower final charges (dashed black and grey lines), a difference that is only seen at lower black hole masses where the Hawking emission is thermal.

{Let's consider the case of a black hole such as that in the Cygnus X-1 binary.  Its mass is estimated at $\sim 21M_\odot$ and its spin is thought to be close to maximal $a>0.9985$ \cite{Zhao}.  The high spin is believed to be due at least in part to accretion from the companion.  It is then possible that black holes in binaries where the star is no longer feeding the black hole are characterized by such high spins.  Such a black hole would be able to acquire a charge from a virial density plasma that ranges between $Q\in [10^{19},10^{25}]$e for spins between $a=0$ and $a=0.999999$.  }

The bottom panel shows that when the plasma density is larger, the accretion increases and allows a larger black hole equilibrium charge. An increase in the baryon density of $6$ orders of magnitude with respect to the average virial density, produces an increase in the charge of almost one order of magnitude (see also the dot-dashed lines in the upper panel).  %This needs to be taken into account for calculations of, for instance, magnetic fields produced by black hole charges (as long as the black hole mass is below the longevous limit).
{The dependence of the charge with redshift can also be inferred from this panel, as the virial density scales with the critical one.  Between $z=0$ and $z=20$ the increase in virial density is of about 3 orders of magnitude, which turns into a charge increase of about $50$ percent.}

\section{Net charge of galaxies}
\label{sec:halocharge}

We now explore galaxy wide consequences of black holes containing electric charge.  We will treat separately the case of primordial black holes which are able to preserve the charge with which they are formed, that A23 propose as being the case for maximally rotating primordial black holes (see for instance \cite{Mirbabayi20} for a discussion of spin of primordial black holes) which could contain positive and negative charges of the  amplitudes shown by the circle and cyan lines of the top panel of Fig. \ref{fig:charge}.  In principle these black holes do not exchange charge with the ionised plasma, leaving the latter perfectly neutral in a global sense.
On the other hand, non-maximally rotating black holes of primordial or astrophysical origin, could extract a negative charge from the plasma, leaving the latter positively charged, but still leaving each individual galaxy perfectly neutral.  

\begin{figure}
    \centering
    \includegraphics[scale=.5]{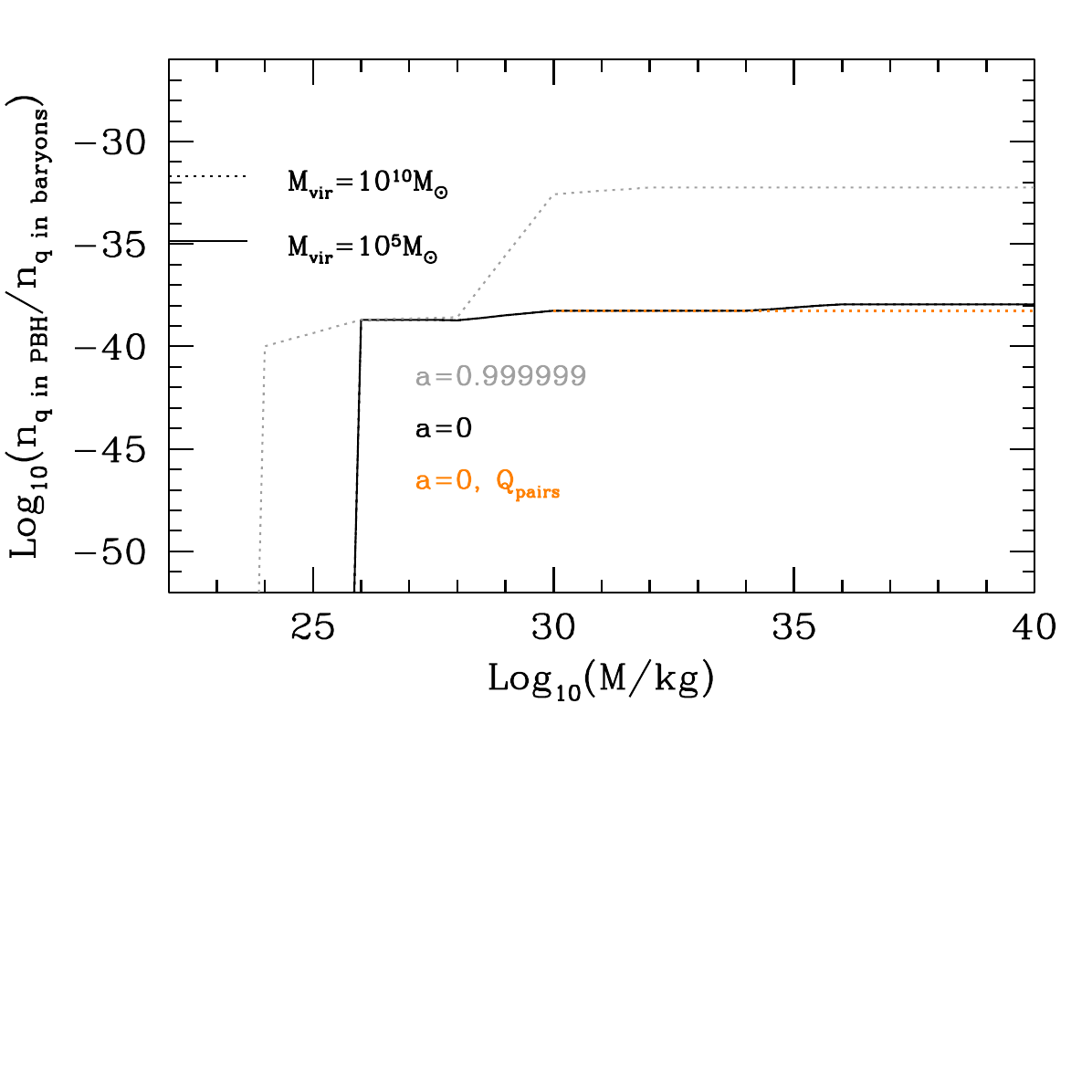}
    \vskip -3.5cm
    
    \caption{Ratio of charges stored in {primordial} black holes, {here assumed to make up all of the dark matter,} and in the {galaxy} plasma, as a function of black hole mass, for different {virial} masses, black hole spins and for $Q_{pairs}$, as shown in the key.}
    \label{fig:halocharge2}
\end{figure}

If black holes are extracting charge from the plasma, this could be at odds with the typically assumed global neutrality of the plasma in galaxies.  Fig. \ref{fig:halocharge2} shows the ratio of charges in {primordial} black holes taken from the plasma and the total number of protons in the plasma, as a function of black hole mass, for different black hole spins and {virial} masses (black and grey, solid and dotted lines as indicated in the key), and also for the case when black holes are seen from afar to hold only $Q^{\rm pairs}$ (orange).  As can be seen, even in the most extreme cases this ratio is below $10^{-30}$, which is a very low level of non-neutrality that does not preclude assuming neutrality for the plasma.  {Stellar black holes would imprint an even lower imbalance of charges in the plasma since they can only make a subpercent fraction of the virial mass.}  {In this simple view where black holes take away the charge from the plasma, an acceptable imbalance of $10^{-30}$  allows to assume neutrality but one should bear in mind that the plasma would have a small, in relative terms, non-zero net charge.  For a typical Milky Way like galaxy the plasma charge would correspond roughly to $10^{32}e$ (with an equal charge of oposite sign in black holes). }

If the {galaxy} {as a whole, including dark matter, stars, black holes and its cold and hot gas,} is neutral it is interesting to wonder if it would be possible for longevous charges in black holes (or simply $Q^{\rm pairs}$), to produce a  non-zero net magnetic {or electric} field {within the galaxy by partially displacing charge and producing local charge imbalances}.  Considering that black holes are collisionless, whereas the plasma is pressure supported, one can expect there to be a net magnetic {or electric} field simply due to the different density profiles of collisionless and collisional fluids.  {We present an example with primordial black holes making up $100\%$ of the dark matter, which follow} a Navarro, Frenk and White \cite{NFW} profile (NFW), with a concentration vs. {virial} mass and redshift taken from \cite{ludlow}.  We assume a core profile for the baryons \cite{Lacey16} which are expected to form a singular isothermal sphere \cite{springel}.  We calculate the
{net charge profile as a function of distance from the galaxy center.  }  {Notice that the net charge at distances larger than the virial radius of the galaxy would be zero.}

\begin{figure}
    \centering
    \includegraphics[scale=.5]{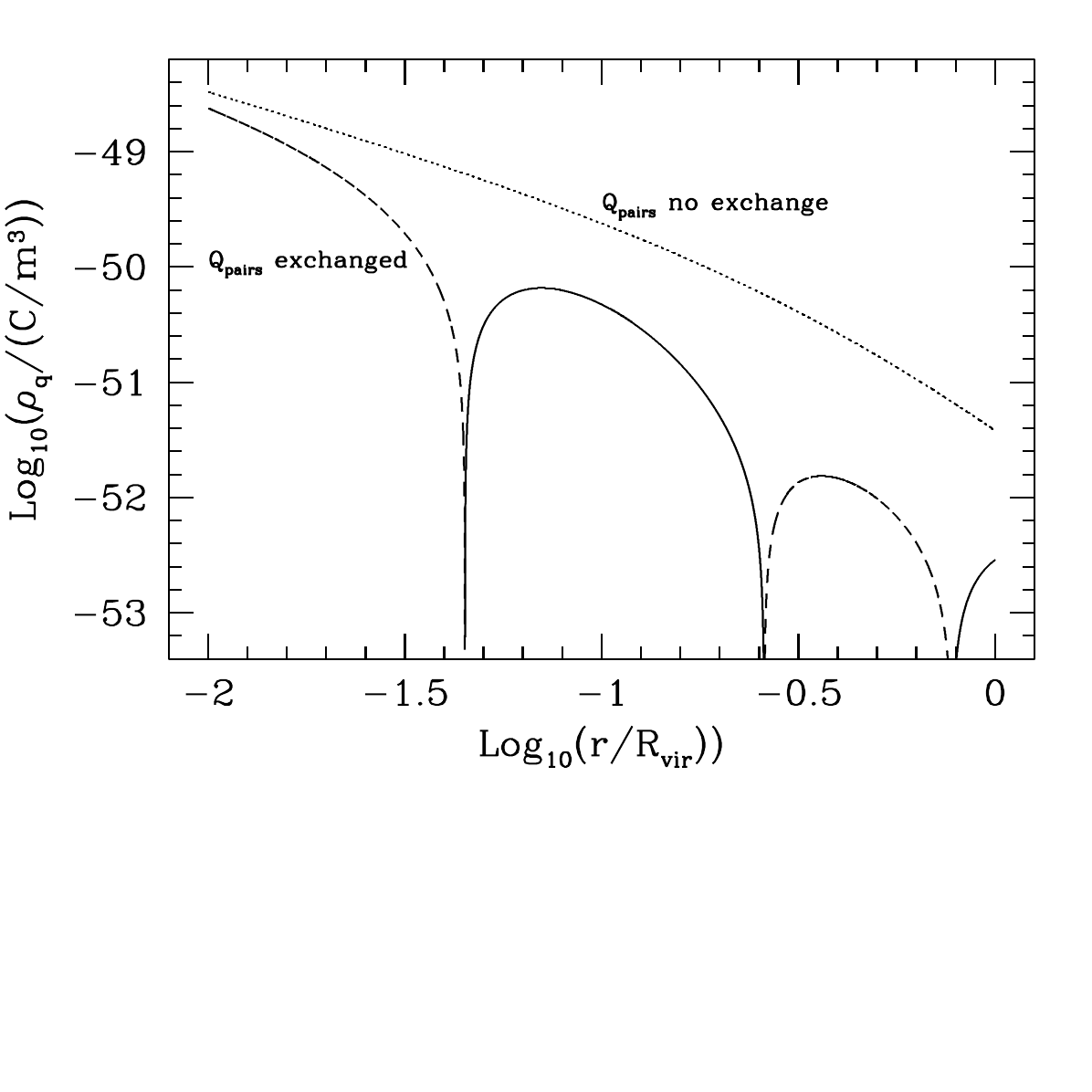}
    \vskip -3cm \caption{ {Charge density profile for a Milky Way sized halo progenitor at $z=20$ with black holes charged with $Q_{\rm pairs}$.  The case where the galaxy is overall electrically neutral is shown as solid lines for positive charges and dashed for negative charges).  The case where the galaxy has a net charge is shown as dotted lines.}  }
    \label{fig:profile}
\end{figure}

%magnetic moment that each fluid would produce, as a function of distance from the halo center, 
%\begin{equation}
 %   m(r)=\left|\frac{1}{2}\int_V \vec r \times (\rho_Q \vec v) dV \right|,
%\end{equation}
%where we assume a specific angular momentum that is equal for ionised %baryons and black holes, and simplify the calculation taking purely %circular motions.  
%Here we can ignore the actual value of the charges and assume that each fluid contains the same total charge but of opposite sign.% (i.e. magnetic moments pointing in exact opposite directions, with different amplitudes as a function of distance to the center of the halo/galaxy).  The resulting magnetic moments 
{We show the resulting charge profiles for a Milky Way progenitor populated by primordial black holes with charge $Q_{\rm pairs}$ at $z=20$ in Fig. \ref{fig:profile}.  We concentrate here on high redshifts since we will later compare these charges to those required to produce a seed for the cosmic magnetic field.  For $z=0$ the resulting charges are only $\sim 30$ percent smaller. We show two cases, one where there is exchange between primordial black holes and the plasma, shown as the solid and dashed line for positive and negative charge density, respectively.  For comparison, we also show the case for no exchange of charges with the plasma, i.e. a hypothetical case of maximally spinning black holes with $Q_{\rm pairs}$, where the amplitude of the charge is larger by about a factor of $10$ throughout.  }  {Notice that at larger distances from the galaxy centre the charge rapidly tends to zero, and more rapidly so in the case of charge exchange between the plasma and black holes.} %We show this ratio for two redshifts for a final $z=0$ halo mass similar to that of the Milky Way, and in both cases we find that the amplitude of the magnetic moment for each fluid is different, except  slightly above the virial radius $r_{\rm vir}$ (equal amplitudes imply zero net magnetic field as the magnetic moments of both fluids are assumed to be directed in opposite directions).  %The sudden drop of some of the lines at low radii reflects that in the inner regions of haloes  the plasma has cooled down and is no longer ionised.  

%This result allows us to simply report the magnetic moment of black holes with NFW profiles as an order of magnitude estimate of the net magnetic moment of the halo in the following Subsection.  If needed, the reader can also obtain more accurate estimates as a function of distance to the halo/galaxy centre with the ratios shown in this figure.

{In light of these results, in the following subsections we explore the total galaxy charge in black holes and take these as an order of magnitude estimate of the net charge of galaxies at roughly their virial radius. }


\subsection{Total galaxy charge from maximally spinning primordial black holes}

In this subsection we will assume that maximally spinning primordial black holes\footnote{{The analysis of this subsection is not applied to stellar remnant black holes because is not believed at present that black holes of stellar origin can reach the maximal spin   \cite{thorne74}}.} (PBH) of equal mass conform all of the dark matter.  If primordial black holes form from the initial fluctuation field left over from Inflation (see for instance \cite{Khlopov_2010}), then they can be assumed to be randomly distributed at late times when  {galaxy} formation starts. In this case, one immediate question is whether the overdense patch containing the black holes and baryons that will collapse to form a {galaxy}, is electrically neutral.  This would indeed be the case if the charged black holes were able to conform a plasma.

\begin{figure}
    \centering
    \includegraphics[scale=.5]{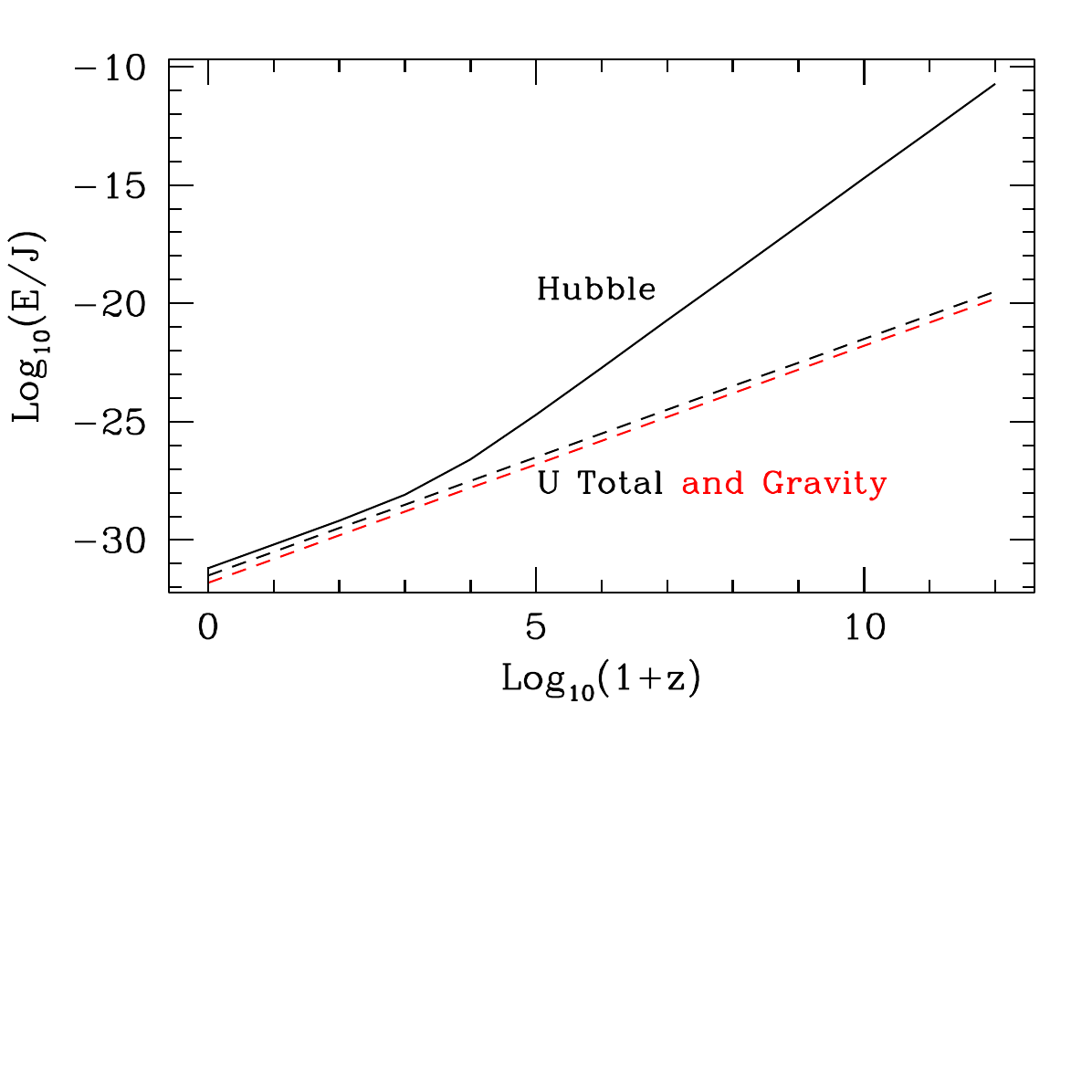}
    \vskip -3.5cm
    \caption{Comparison between the available total (Coulomb and gravitational) and gravitational energy (black and red dashed lines, respectively) exherted by Planck mass black holes with maximal RN charge, that lie within overdense patches that will collapse to form {galaxies}, and the energy equivalent of the expansion of the universe on the same scales (black solid), at a scale of the mean separation between black holes.  Since Coulomb and gravity cannot overcome the expansion, it can be safely assumed that charged, maximally spinning black holes with positive and negative initial charges, cannot relax into a plasma and their charges will remain unrelaxed and essentially randomly distributed in space down to the present day.}
    \label{fig:poissonhalo}
\end{figure}

In order to test this we make a simple calculation of the potential Gravitational and Coulomb energy exerted between PBHs of Planck mass holding the extremal RN charge, which are the ones that present the highest Coulomb to gravitational force ratios, and would therefore present the highest chances of forming a plasma.  To do the calculation we take into account the mean physical distance between these PBHs as a function of redshift, and compare the Coulomb and gravitational potentials at these separations with the required kinetic energy that the PBHs would need to acquire in order to offset the expansion of the Universe on the same scales.  This is shown in Fig. \ref{fig:poissonhalo} where regardless of the redshift, the Coulomb and gravitational potentials are at least a factor of a few below the needed levels to overcome expansion.

Armed with this result, we can postulate that regardless of origin and mass, primordial black holes will not be able to relax into a plasma configuration, and therefore {galaxies} will be endowed with a net charge arising from the Poisson imbalance of primordial black holes with positive and negative charges, $Q(M)$, within the initial overdense patch from which the {galaxy} collapsed.  We will refer to this {galaxy} charge as the galaxy initial Poisson charge,
\begin{equation}
    Q^i_p(M_{\rm vir})\sim Q(M) \sqrt{M_{\rm vir}/M}.\label{Eq:qphalo}
\end{equation}
The resulting galaxy charges can be of both positive and negative sign, and are shown as red symbols and cyan lines in Fig. \ref{fig:halocharge}.

\subsection{Net charges in galaxies with non-maximally spinning black holes}

\begin{figure}
    \centering
    \includegraphics[scale=.5]{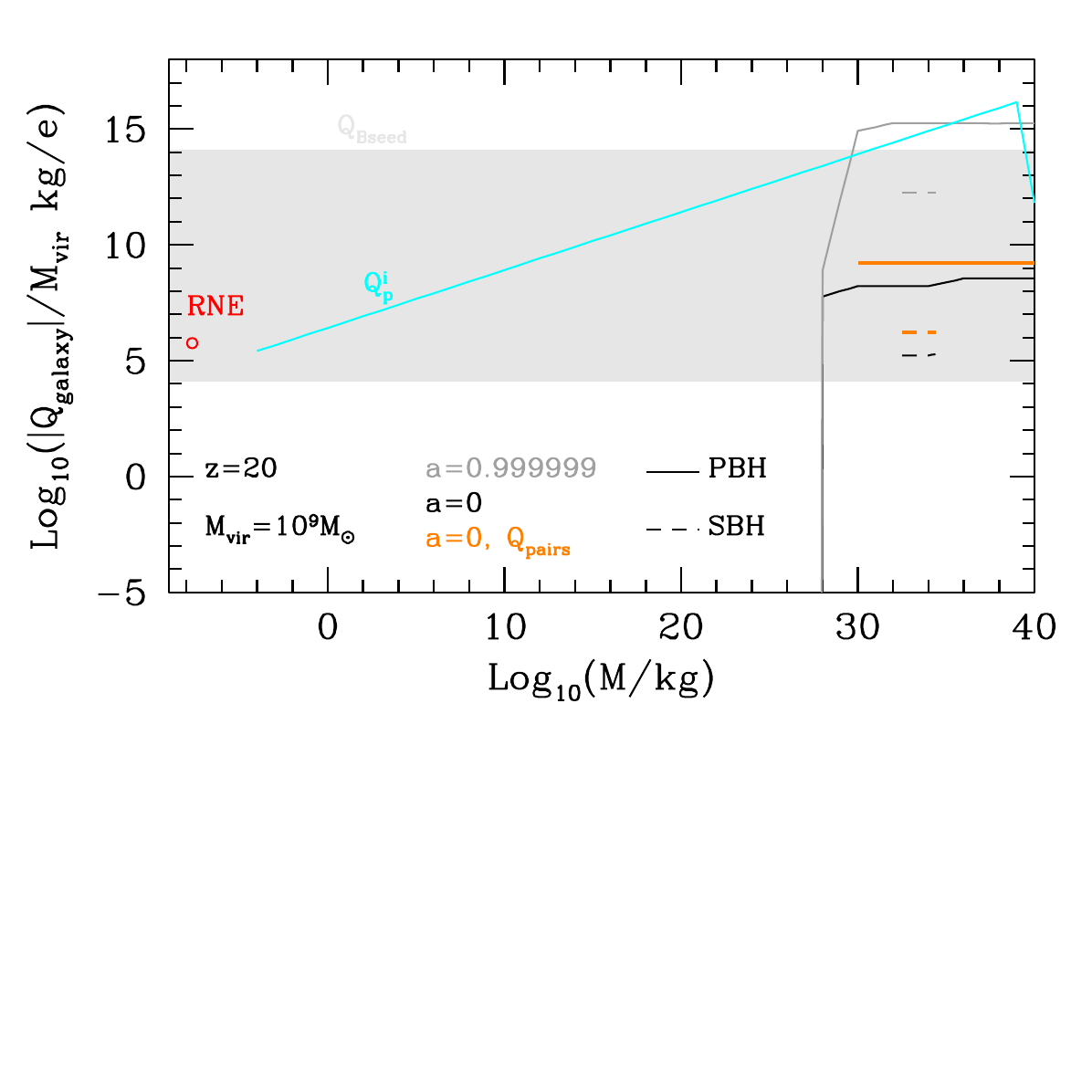}
    \vskip -3.5cm
    \caption{{Total galaxy} charge per mass in primordial {(solid) or astrophysical black holes (dashed lines) at $z=20$}, as a function of black hole mass. Red and cyan show the net {galaxy} charge due to Poisson noise in the number of positive and negative maximally spinning primordial black holes within the overdense patch that gave rise to the formation of the {galaxy}.  Black and grey lines show the net charge in non-maximally spinning black holes alone, considering  two different values for the dimensionless spin parameter.  The orange line shows the {galaxy} charge in black holes considering the pair production black hole charge $Q_{pairs}$. 
    {The shaded light grey region shows the total galaxy charge needed to produce a magnetic field seed of $10^{-30\pm5}G$ for this virial mass.} }
    \label{fig:halocharge}
\end{figure}

In the case of non-maximally spinning black holes, {of either stellar or primordial origin,} their  charge is acquired from the plasma, which makes the galaxy electrically neutral.  Excluding any high energy processes that could expel baryon charges away from the galaxy, or any intrinsic sources of charge imbalance within galaxies \cite{caprini2005,Goolsby2015}, the total charge in black holes within a galaxy equals that in the plasma or the baryons in general within that same {galaxy} with the opposite sign.  In this case the total {galaxy} charge in black holes is simply the product of the number of black holes in the {galaxy} and of their charge, and is shown as black and grey lines in Fig. \ref{fig:halocharge} for the virial mass of a Milky Way progenitor at $z=20$.  We show results for different black hole dimensionless spin parameters, and do so for primordial and stellar ones, as shown in the key.  As it can be seen, the charge is rather constant since when $Q>0$ the scaling of the black hole charge is $\propto M$.  The {galaxy} charge is higher for higher black hole spins.  %At low black hole masses the charge is lower than the total halo charge for maximally spinning primordial black holes (cyan lines and red symbols), while they

The amplitude of the charge for PBHs with $a<1$ can be similar to that obtained with maximally spinning PBHs, depending on the black hole mass and spin.  The orange dotted line shows the charge in black holes considering a charge of $Q^{\rm pairs}$ which, as postulated above, is the minimum charge that black holes with $M\gtrsim 10^{22}kg$ (a value that depends on the plasma density) could be seen to hold from afar, even if their equilibrium charge was higher. 

{The charge corresponding to astrophysical black holes (SBH) assumes these live in the stellar halo of a galaxy so as to allow them to acquire a charge via their interaction with the ionised plasma within the galaxy}.  In this case we adopt the measured logarithmic slope of the density profile of halo stars in the Galaxy from Slater et al. \cite{Slater2016} of $-3.5$.  Again we assume that the amount of charge is the same but of opposite sign for astrophysical black holes and the plasma.  Even though the spatial extent of the stellar halo is in principle more confined than that of the hot plasma or the dark matter, a large fraction of the stellar (and stellar products) content of the halo are thought to have formed ex situ, i.e. outside the galaxy, and to have been deposited in the halo as a result of mergers (see for instance, Pilepich \& Madau \cite{Pilepich2015}).  Therefore, stellar halo  black holes could have acquired a charge during their infall process as they came into contact with the ionised plasma of their new host galaxy.  This justifies assuming the charge to have the full extent of the distribution of gas and stellar black holes.  In this case the charge density profile has a steeper drop than the isothermal ionised gas of the galaxy.  

{The resulting amplitude of the charge density in stellar black holes in the galaxy is about three orders of magnitude lower  as the case of primordial black holes because of the fraction of mass in stellar black holes.  The latter is obtained assuming the star formation efficiency of \cite{Behroozi13} and  that stars with $m_{\rm stellar}>20M_\odot$ form black holes at the end of their lives taking a Salpeter  initial stellar mass function \cite{salpeter}.  The resulting charges are shown by the dashed lines in Fig. \ref{fig:profile}, over the narrow range of masses that SBH are expected to cover for different black hole spins.  The grey dashed line is the extreme case where stellar black holes  include a sizeable population of high spin stellar remnants such as the one in Cygnus X-1, where the charge could be higher by up to six orders of magnitude.}

%We also explore t

{In these calculations we have simply ignored the multiple phase nature of the gas in galaxies, since we concentrated on very high redshifts where the virialization is recent and little cooling could have taken place.  This assumption is likely more inaccurate for stellar black holes as these require star formation, and therefore cooling, to have already taken place.  We will improve these assumptions applying these calculations on hydro simulations in Colazo et al. (in prep.).}
%The ratio of magnetic moments between plasma and SBH shown as a grey line in Fig. \ref{fig:profile} indicates that it is indeed possible for SBHs to produce a  non-zero net magnetic moment within galaxies, except at roughly $1/10$ of the virial radius.

\subsection{Required charges to provide cosmic magnetic seeds}
\label{ssec:qseed}

{Even though there are several possible viable options to provide the required primordial seeds of cosmic magnetic fields \cite{Grasso01,Dar05,Quashnock89,Banerjee03,Durrer06,Saga15,zhou,Xu2008}, charged black holes within a recently collapsed early galaxy could also provide a dipolar magnetic field at very early times, contributing to the seed magnetic fields at the time of the onset of astrophysical dynamos.  }
{One can compare the black hole charges to the ones needed to produce a magnetic field at early times, $z\sim 20$, with the required amplitude to explain present-day galaxy-wide cosmic magnetic fields.  {Even though there are proposals that produce higher amplitude seeds \cite{Xu2008}, we adopt the minimum amplitude that is needed according to Davis et al. \cite{Davis1999}, of }  $B_{\rm seed}\sim10^{-30\pm5}G$, {where the large uncertainty is due to the difficulty assessing the dynamo e-folding timescales. } Astrophysical processes that take place afterwards are able to increase the magnetic field by more than $20$ orders of magnitude, { reaching the expected $B_{\rm fil}\sim10^{-9}G$ in cosmic filaments or $B_{\rm gal}\sim10^{-9}G$ in galactic halos \cite{Vazza2021}}.}

{We make a simple calculation assuming that all of the dark matter is made of black holes and that these are distributed in an NFW profile with dimensionless {galaxy} spin parameter $\lambda'=J_{\rm vir}/(M_{\rm vir}V_{\rm vir})$, where $J_{\rm vir}$ and $V_{\rm vir}$ are the angular momentum and virial velocity, respectively \cite{maccio07}, %assume a scaling for the halo concentration of the form $\log_{10} c=1.02 -0.109 (\log_{10}(M_{\rm vir}/M_{\odot})-12)$ from \cite{ludlow}, 
with which the amplitude of the magnetic field at the virial radius, as a function of the mass of the black holes that make up the {galaxy}, reads,
\begin{equation}
    B(r_{\rm vir})=\frac{\mu_0}{4\pi}\frac{m(r_{\rm vir})}{r_{\rm vir}^3}=\frac{\mu_0}{4\pi}\frac{1}{2}\frac{Q}{M}\lambda' \sqrt{2}\frac{M_{\rm vir}V_{\rm vir}}{r_{\rm vir}^2}\label{eq:B}
\end{equation}
%\begin{equation}
%    B(r_{\rm 1/2})=\frac{\mu_0}{4\pi}\frac{m(r_{\rm 1/2})}{r_{\rm 1/2}^3}=\frac{\mu_0}{4\pi}\frac{1}{2}\frac{Q}{M}\lambda' \sqrt{2}\frac{M_{\rm gas}V_{\rm vir}}{r_{\rm 1/2}^2}
%\end{equation}
where  $\mu_0$ is the magnetic permeability of the vacuum, $m$ is the magnetic moment, $r_{\rm vir}$ is the virial radius, and we have assumed $\log_{10} \lambda'=-1.5$ \cite{maccio07}.  The required charge per mass can then be written as
\begin{equation}
    Q/M=\frac{8 \pi B_{\rm seed} r_{\rm vir}^2}{\sqrt2 \mu_0 \lambda' M_{\rm vir}V_{\rm vir}},
\end{equation}
{for black holes with equal sign charge (i.e. for comparison to  acquired charges only).}
%\begin{equation}
%    Q/M=\frac{8 \pi B_{\rm seed} r_{\rm vir}^2}{\sqrt2 \mu_0 \lambda' M_{\rm gas}V_{\rm vir}}\sim \frac{e}{10^{-18}M_\odot}.
%\end{equation}
The light grey {shaded region in  Fig. \ref{fig:halocharge} shows the range that would produce the seed (within its estimated errors), which encompasses the charges obtained from the different types of black holes (primordial or stellar, maximally spinning or not) shown in the figure. } 

Notice though, that this simple calculation does not take into account whether black holes with these charges would be able to produce a galaxy wide non-zero charge and magnetic fields {since it does not take into account that the charges in black holes could come from the galactic plasma; instead of a {galaxy} net charge, there could be a non-zero galaxy charge that depends on the distance to the galaxy center as shown in Fig. \ref{fig:profile}.  Our aim with this comparison is to point out the interesting coincidence between the required charge for magnetic seeds and the charges inferred for {galaxies} with charged black holes. We leave an in depth analysis of these details for a forthcoming work (Colazo et al., in prep)}.

\section{Discussion}
\label{sec:discussion}

In this section we explore in more detail some caveats that could affect our conclusions.

We start looking at the effect that accretion discs could have on the charge of the black holes, which we have ignored up to this point.  We expect accretion discs to be able to form around black holes when the Bondi radius \cite{Bondi_1952} $r_B=2GM/\tilde{v}^2$ (where $\tilde{v}$ is the sum in quadrature of the sound speed and the virial velocity)  exceeds the Debye length of the plasma.  Since the main conclusions here are centred on high redshift {galaxies}, we compare these two quantities for $z=20$ and $M_{\rm vir}\sim 10^{10}M_\odot$.  In this case, assuming a fully ionised plasma at the virial temperature, we find that black holes with $M>10^{22}kg$ would be able to form accretion discs.  If the radiation field within the {galaxy}, coming either from the accretion onto a budding supermassive black hole or from intense bursts of initial star formation, is able to increase the plasma temperature, this would increase the threshold mass for accretion discs.  For a temperature $1000$ times higher \cite{ohmura}, the minimum mass becomes $M\sim10^{27}kg$, which is below the solar mass.  Therefore, it is quite plausible that the charges of BHs below this threshold mass are free of the effect of an accretion disc.  %However, masses around the sub-solar mass and higher could be affected by an accretion disc.  

We have also mentioned earlier that the charge of BHs could be neutralised by a trailing group of protons with equal charge.  We remind the reader that we  consider that BHs and the plasma are dynamically independent.  Since BHs with $M>10^{22}-10^{27}$kg and higher (depending on the plasma density and redshift) could form an accretion disc, they would in principle be able to take along charges of opposite sign after them simply because these would be confined within the Bondi mass bound to the BH. %This mass would be extracted from the plasma since we assume that accretion occurs only when the Debye length of the plasma is smaller than the Bondi radius.  
This could indeed neutralise the charge as seen by a distant observer.  However, if charges were acquired  faster  than  the binding time of the Bondi mass, then the gravitational field of the black hole would not have enough time to take along the compensating protons; they would be left behind trapped in the plasma.   We take the free fall timescale for the Bondi mass for BHs in a plasma with $\rho_{vir}$, $t_B=\sqrt{r_B^3/(2GM)}$, as the binding timescale for the Bondi mass and show its dependence on black hole mass as the violet dotted line in Fig. \ref{fig:timescales} (the line starts at the minimum BH mass that can form an accretion disc for a surrounding plasma at $T_{\rm vir}$).  As it can be seen, $Q_{\rm pairs}$ can be acquired much faster for $M>10^{28}kg$, suggesting that no compensating charges will be carried along by the BH for this charge.  Longevous charges below $Q_{\rm pairs}$ could also be affected, as the Bondi timescale is shorter than the charging time for lower mass BHs.

One important possibility that could discharge black holes completely is that the accretion disc around stellar or primordial black holes were energetic enough to produce positron-electron pairs; the emitted positrons would be able to discharge the black hole. Laurent et al. \cite{Laurent2018} show that the emission in some cases is as high as $>10^{35}$ positrons per second.  However,  Sarkar \& Chattopadhyay \cite{Sarkar2020} show that the positron emission is non-zero only when the Eddington normalised accretion rate is $\dot M/\dot M_{Edd}>0.8$, where $M_{Edd}=4\pi G M m_p/(c \sigma_T)$, with $\sigma_T$ the Thompson scattering cross-section.  

We calculate the Bondi-Hoyle-Lyttleton \cite{Bondi-Hoyle_1994, Hoyle_Lyttleton_1939, Bondi_1952} accretion rates,
\begin{eqnarray}
\dot{M}=4 \pi r_{B}^{2} \tilde{v} \rho=\frac{4 \pi G^{2} M^{2} n \mu m_{p}}{\tilde{v}^{3}}
\label{eq:M_dot}
\end{eqnarray} 
where $\mu$ is the mean molecular weight and $n$ is the gas number density. We find that all black holes with  $M<10^{34}kg$ (i.e. $\sim 10^4M_\odot$)  in our calculations accrete well below $10$ percent of the Eddington limit as long as the baryon density around black holes is $<10^6 \Omega_b \rho_{\rm vir}$ (the density of molecular clouds, for example), indicating that the charges and seed magnetic fields estimated in this paper should not be affected by positron emission in accretion discs, i.e., it is reasonable to assume that accretion discs do not affect the charges of BHs with $M\lesssim10^4 M_\odot$.  The exception would be stellar black holes in high mass X-ray binaries, but only while they acrete from their companion.  Once this process stops, their accretion rates fall below the threshold for electron-positron spontaneous emission.  On the other hand, supermassive black holes (even those orbiting within the galaxy, coming from merged satellites) would not be expected to hold a non-zero electrical charge.

We also point out that our calculation of {charge profiles} %magnetic field seeds 
assumes that the black holes hold charge throughout the spatial extent of a galaxy.  We have purposely ignored that the gas in a galaxy {more complicated than our simple two-phase model} with a hot ionised phase and a cold, neutral one.  We have done so because we are assuming that it is enough for black holes to at least come into contact with the ionised phase at some point, and this is justified by concentrating on long-lived charges, on charges below the Gibbons limit, or on a fraction of the black hole charge corresponding to $Q_{pairs}$, given the fast rates of charge acquisition by black holes for all these cases.  {However, the effect of the BH charges on the frontier between neutral and ionised gas within galaxies needs to be modelled further.  Our present calculations should be robust at least in regions lying within the ionised plasma, which can still imprint non-zero magnetic fields due to the different local steepness of the black hole and gas density profiles.}

%We end by noticing also that we have shown quite a broad range of halo masses in our estimates of magnetic field seeds and that%, for black hole masses $M>10^{11}kg$, 
%the increase in magnetic field amplitude is smaller than that in halo mass.  This indicates that our results are not too sensitive to the halo mass and, in turn, to the redshift where the seed is calculated.  The magnetic field seed for a Milky Way-like progenitor would only grow by roughly  three orders of magnitude from $z=15$ to $z=0$, whereas astrophysical processes can increase the amplitude of the magnetic field by tens of orders of magnitude \cite{Davis1999}.  This shows that charged black holes, even at the Gibbons limit, provide seeds of roughly the adequate amplitude independently of the exact values adopted for redshift and halo mass.

\section{Conclusions}
\label{sec:conclusions}

In this paper we extend the formalism presented in Araya et al. (2023, \cite{Araya23}, A23) estimating the charge that non-maximally spinning black holes embedded in a baryonic plasma are able to attain.   In particular we study the rates of Hawking athermal charge emission, and of charge accretion from the surrounding ionised plasma, which allows us to estimate an equilibrium charge and its dependence on the black hole spin for dimensionless spin parameters $a<1$.  We find that black holes either surrounded by high density plasma or with high spins, are indeed able to surpass the Gibbons 1975 \cite{Gibbons75} limit of discharge by pair particle production $Q_{pairs}$ for black holes of $M>10^{22}$kg. {This is mostly due to the limiting cross sections for charge carriers which on the lower limit have a radius equal to that of the black hole, and on the upper limit the Debye length; in a galactic plasma both are of the same order of magnitude for these black hole masses.  We also find that lower mass black holes would not hold a charge for a significant fraction of the time.}

We also note the distant observer would only be able to infer at most charges that are longevous or a charge of $Q_{pairs}$ for a higher equilibrium charge, since this is the only charge that black holes would be able to effectively  spatially displace from the baryon plasma.  This formalism can be applied indistinctively to primordial and astrophysical black holes.

We  explore the net charges %and magnetic moments 
of galaxies and the dark matter haloes that host them considering {first, the case where } primordial black holes make up all of the dark matter {and, second, a case where we consider only stellar black holes}.

In the first case, if the black holes are maximally spinning and retain the charge they formed with, or even if their spin is non-maximal and the charge is acquired from the plasma, then {the total galaxy charge ranges between $|Q_{\rm galaxy}|/M_{\rm vir}=10^6$ to $10^{16}$kg$/e$, depending on black hole mass, spin, and whether one considers the total black hole charge or only $Q_{\rm pairs}$, which is expected to be a charge that can be displaced on galaxy-wide scales. }%their charge combined with the typical angular momentum of galaxies and dark matter haloes %, which would hold a net charge, or displaced charges by black holes, 
%is enough to produce magnetic fields of $\sim 10^{-30}$Gauss for black hole masses $M>10^{26}$kg (even for $M>10^{22}kg$ for high density plasmas).  Notice that the current mass windows where primordial black holes can make up all of the dark matter lie roughly at $M\sim 10^{10}$ and $10^{30}$kg, i.e. asteroid and solar masses, respectively \cite{Sureda:2020,Carr:2020b}, presenting primordial black holes of solar mass with any spin, and maximally spinning ones of asteroid mass, as possible candidates to produce the magnetic field seeds regardless  of the existence of magnetic monopole charges within them (see also \cite{araya}). 

%Another interesting result is the following.  
{Regarding the second case, we first note that} current numerical simulations of galaxy formation, as well as observational evidence, suggest that the stellar haloes of galaxies contain high fractions of ex-situ formed stars \cite{Slater2016} and, consequently, black holes.  
{Adopting only the fraction of mass in stellar black holes in galaxy stellar haloes, the total estimated galaxy charge lies at $|Q_{\rm galaxy}|/M_{\rm vir}=10^5$kg$/e$.}

{We also perform a simple calculation pointing out that a rotating galaxy with a non-zero charge profile would produce a net magnetic field.  We calculate the galaxy charge that is required in order to produce the seed magnetic field at very high redshifts, right after the collapse of the progenitor halo of galaxies.  Taking into account the uncertainties in the required field seeds of $B_{\rm seed}\sim10^{-30\pm5}$G \cite{Davis1999}, this charge lies in the range $|Q_{\rm galaxy}|/M_{\rm vir}=10^4$ to $10^{14}$kg$/e$.  Even though this calculation is simplistic and  there are several proposals to produce the seed in the literature \cite{Grasso01,Dar05,Quashnock89,Banerjee03,Durrer06,Saga15,zhou} {including ones that produce quite higher initial seeds (e.g. \cite{Xu2008}),} it shows that the charges from primordial or even stellar black holes are indeed another possible seed generating mechanism.  In this case the seed would not be primordial as produced by other mechanisms, but it could be there in place by the time galaxy formation begins, having a distinctive spatial distribution that, in addition to observational features from other magnetogenesis scenarios, could be tested observationally in the near future}. 
%Considering the leeway in the estimated magnetic field seeds needed to explain the observed ones today \cite{Davis1999}, we find that even astrophysical black holes in the stellar haloes of galaxies could provide the magnetic field seeds.  To do so these black holes would indeed need to be accreted from merged galaxy satellites so that they would have been at some point embedded in the hot, ionised gas halo surrounding galaxies to acquire a charge; even if they are able to accrete just the Gibbons pair production charge $Q_{pairs}$.  This charge is quite low, producing a Coulomb potential some $\sim20$ orders of magnitude lower than the gravitational one regardless of black hole mass.  This is probably the reason why it is commonly regarded that the black hole charge can be safely ignored.  However, we have shown here that even such low level of stellar black hole charge is able to provide seeds of magnetic field %of the necessary amplitude.
% the resulting seeds are quite 
%close to the required values to explain the cosmic magnetic fields observed today.  

\acknowledgments

We thank Juan Magaña, Joaquín Armijo, Marcelo Rubio, Anna Durrant and Patricio Colazo for helpful comments and discussions.  NDP acknowledges support from a RAICES, a RAICES-Federal, and PICT-2021-I-A-00700 grants from the Ministerio de Ciencia, Tecnología e Innovación, Argentina.  The work of IJA is funded by ANID FONDECYT grants No.~11230419 and~1231133, and by ANID Becas Chile grant No.~74220042. IJA also acknowledges funding by ANID, REC Convocatoria Nacional Subvenci\'on a Instalaci\'on en la Academia Convocatoria A\~no 2020, Folio PAI77200097. FAS thanks support by grants PIP 11220130100365CO,
PICT-2016-4174, PICT-2021-GRF-00719 and Consolidar-2018-2020, from CONICET, FONCyT (Argentina) and SECyT-UNC.
IJA is grateful to Andrei Parnachev and the School of Mathematics at Trinity College Dublin for their hospitality. 

\bibliographystyle{unsrt}
\bibliography{references}
\end{document}